\documentclass[aps,prb,twocolumn,amsmath,amssymb,showpacs,superscriptaddress,notitlepage,longbibliography,floatfix]{revtex4-1}
\usepackage{array}
\usepackage{float}
\usepackage{graphicx}
\usepackage{subfigure}
\usepackage{dcolumn}
\usepackage{bm}
\usepackage{amsfonts}
\usepackage{mathrsfs}
\usepackage{amssymb}
\usepackage{amsmath}
\usepackage{color}
\usepackage{braket}
\usepackage{appendix}
\usepackage{chngcntr}
\usepackage{stmaryrd}
\usepackage{xcolor}
\usepackage[colorlinks=true,breaklinks=true,linkcolor=blue,anchorcolor=blue,citecolor=blue,urlcolor=blue]{hyperref}
\usepackage{booktabs}
\usepackage{multirow}
\usepackage{tcolorbox}
\usepackage{titlesec}
\usepackage{amsthm}
\usepackage{lineno}
\usepackage[T1]{fontenc}

\usepackage{etoolbox}
\makeatletter
\patchcmd{\@makeschapterhead}{\vspace*{50\p@}}{\vspace*{0pt}}{}{}
\renewcommand{\bibsection}
\par
\makeatother

\titleformat{\section}[block]{\normalfont\bfseries\centering}{\thesection.}{0.5em}{}

\begin{document}

\title{Characterizing second-order topological insulators via entanglement topological invariant in two-dimensional systems}

\author{Yu-Long Zhang}
\affiliation{College of Physics, Hebei Normal University, Shijiazhuang 050024, China}
\affiliation{Department of Physics, Shijiazhuang University, Shijiazhuang 050035, China}

\author{Cheng-Ming Miao}
\affiliation{International Center for Quantum Materials, School of Physics, Peking University, Beijing 100871, China}

\author{Qing-Feng Sun}
\email[]{sunqf@pku.edu.cn}
\affiliation{International Center for Quantum Materials, School of Physics, Peking University, Beijing 100871, China}
\affiliation{Hefei National Laboratory, Hefei 230088, China}

\author{Jian-Jun Liu}
\email[]{liujj@mail.hebtu.edu.cn}
\affiliation{College of Physics, Hebei Normal University, Shijiazhuang 050024, China}
\affiliation{Department of Physics, Shijiazhuang University, Shijiazhuang 050035, China}

\author{Ying-Tao Zhang}
\email[]{zhangyt@mail.hebtu.edu.cn}
\affiliation{College of Physics, Hebei Normal University, Shijiazhuang 050024, China}

\date{\today}

\begin{abstract}
	Higher-order topological insulators have attracted significant interest in recent years. However, identifying a universal topological invariant capable of characterizing higher-order topology remains challenging. Here, we propose a entanglement topological invariant designed to characterize second-order topological systems. This entanglement topological invariant captures the entanglement of topological corner states under open boundary conditions by employing a bipartite entanglement entropy method. In several representative models, the entanglement topological invariant assumes a nonzero value exclusively in the presence of second-order topology, with its magnitude exactly matching the number of topologically protected corner states. Consequently, the proposed entanglement topological invariant not only provides a clear criterion for detecting higher-order topology, but also offers a quantitative measure for the related corner states. Our study establishes a universal and  precise method for characterizing higher-order topological phases, opening avenues for their fundamental understanding and future investigations.
\end{abstract}
	
\maketitle
\noindent
\textbf{\large Introduction}\\
Higher-order topological systems have garnered significant attention in recent years due to their potential to extend the traditional bulk-edge correspondence. This framework assumes that a $d$-dimensional $n$-th order topological system hosts edge states in a $d-n$-dimensional subspace \cite{ Song2017,Langbehn2017,Schindler2018,Franca2018,Wang2019,Ezawa2018,Calugaru2019,Trifunovic2019,Khalaf2018,addref1,Wang2025}.
For instance, a two-dimensional first-order topological system exhibits one-dimensional edge states, while a two-dimensional second-order topological system is characterized by zero-dimensional corner states.
First-order topological phases are well-characterized by topological invariants such as Chern numbers \cite{Thouless1982} and $ \mathbb{Z}_{2} $ invariants \cite{KaneMele2005a, KaneMele2005b}. In contrast, identifying a suitable topological invariant for higher-order phases remains more challenging.
Several proposals have been advanced, including the charge quadrupole moment\cite{Benalcazar2017,Benalcazar2017PRB,Ono2019,Li2020}, nested Wilson loops \cite{Benalcazar2017,Benalcazar2017PRB}, and symmetry-based topological invariants, such as the $v$ analogous to $ \mathbb{Z}_{2} $ \cite{Schindler2018a,Costa2021}, among others. However, these invariants often lack universality across different systems, making it difficult to provide a unified description of higher-order topological phases. Additionally, topological corner states can arise from various topological mechanisms\cite{Miao2022, Miao2023, Miao2024, Yang2020}, further complicating the identification of a single, universal invariant.

In addition to topological invariants, topological entanglement entropy provides another powerful method for characterizing topological phases in many-body systems, as initially proposed by Kitaev-Preskill and Levin-Wen \cite{Kitaev2006, Levin2006}.
The entanglement entropy $S$ of a subsystem exhibits a linear relationship with the boundary length $L$, expressed as $S = \alpha L - \gamma$, where $\gamma$ represents the topological entanglement entropy and is nonzero in nontrivial topological phases.
This quantity reflects a long-range entanglement that cannot be disentangled by local operations, thus establishing a direct connection between quantum entanglement and topology.
The utility of entanglement entropy as a tool for topological characterization has been demonstrated in various many-body systems \cite{Amico2008, Calabrese2009, Eisert2010, Fradkin2013, Tsomokos2009, Halasz2012, Halasz2013}, both theoretically \cite{Depenbrock2012, Jiang2012, Isakov2011} and experimentally \cite{Sankar2023, Karamlou2024, Lin2024,Chen2019}.
Beyond entanglement entropy, the entanglement spectrum has also been shown to characterise information of the edge excitations and serve as a sensitive indicator of topological order in many-body systems \cite{Li2008, Chandran2011, Prodan2010,Mao2025}.
In addition, mutual information has been discussed in the context of topological systems, particularly in Chern-Simons theories, where it captures the topological contribution to the total entanglement entropy and reflects long-range quantum correlations between disjoint non-contractible regions on a torus \cite{Wen2016}. These studies, mainly focused on interacting many-body systems, demonstrate that entanglement-based quantities---including entanglement entropy, entanglement spectrum, and mutual information---serve as powerful probes of first-order topological phase.

Furthermore, entanglement-based approaches have been successfully extended to free-fermion systems, where the entanglement entropy and entanglement spectrum provide efficient and reliable probes of topological properties \cite{Fidkowski2010, Alexandradinata2011}. In particular, when the free-fermion system is a first-order topological insulator or superconductor with gapless edge states, the entanglement spectrum often exhibits \(1/2\) modes \cite{Fidkowski2010}. However, this correspondence is not universal. For instance, topological crystalline insulators with inversion \cite{Hughes2011} or $C_n$ rotation \cite{ Fang2013} symmetry  exhibit symmetry-protected \(1/2\) modes in their entanglement spectrum, even without first-order edge states in the energy spectrum. Such
features have been understood as signatures of higher-order topology, underscoring the sensitivity of entanglement spectra to bulk crystalline symmetries and topological structures beyond first-order phases. Despite
these insights, applying entanglement entropy to systematically characterize higher-order topological phases remains challenging. Traditional formulations of entanglement entropy are typically based on periodic boundary conditions, which are inadequate for capturing boundary-localized features inherent in some higher-order topological phases. In particular, some realizations rely on specific boundary engineering that cannot be described within a periodic framework \cite{Miao2022, Miao2023, Miao2024, Yang2020}.
Therefore, extending entanglement entropy calculations to systems with open boundaries may enable the formulation of entanglement topological invariant (ETI) that can effectively capture higher-order topology.

In this paper, we introduce the ETI, denoted as $S^{T}$, to characterize second-order topological insulators, with a particular focus on the corner states entanglement properties in finite-size  free-fermion systems.
This entanglement arises from corner states induced by specific boundary conditions.
We demonstrate the effectiveness of this approach by applying it to a bilayer-coupled Bernevig-Hughes-Zhang (BHZ) model \cite{Bernevig2006}.
Our results show that ETI $S^{T}$ provides robust characterization of higher-order topological insulators: $S^{T}=0$ corresponds to topologically trivial phases, while $S^{T} = N_{0}$ indicates the presence of higher-order topological phases, with the quantized integer $N_{0}$ representing the number of second-order corner states, i.e., the values of $S^{T}$ are identified as correlate with the number of corner states.
This method is not only applicable to second-order topological insulators but also can extend to second-order topological superconductors, highlighting its broad applicability.
\\
\\
\\
\noindent
\textbf{\large Results}\\
\textbf{The definition of the ETI}\\
\begin{figure}
	\centering
	\includegraphics[width=1\columnwidth,clip]{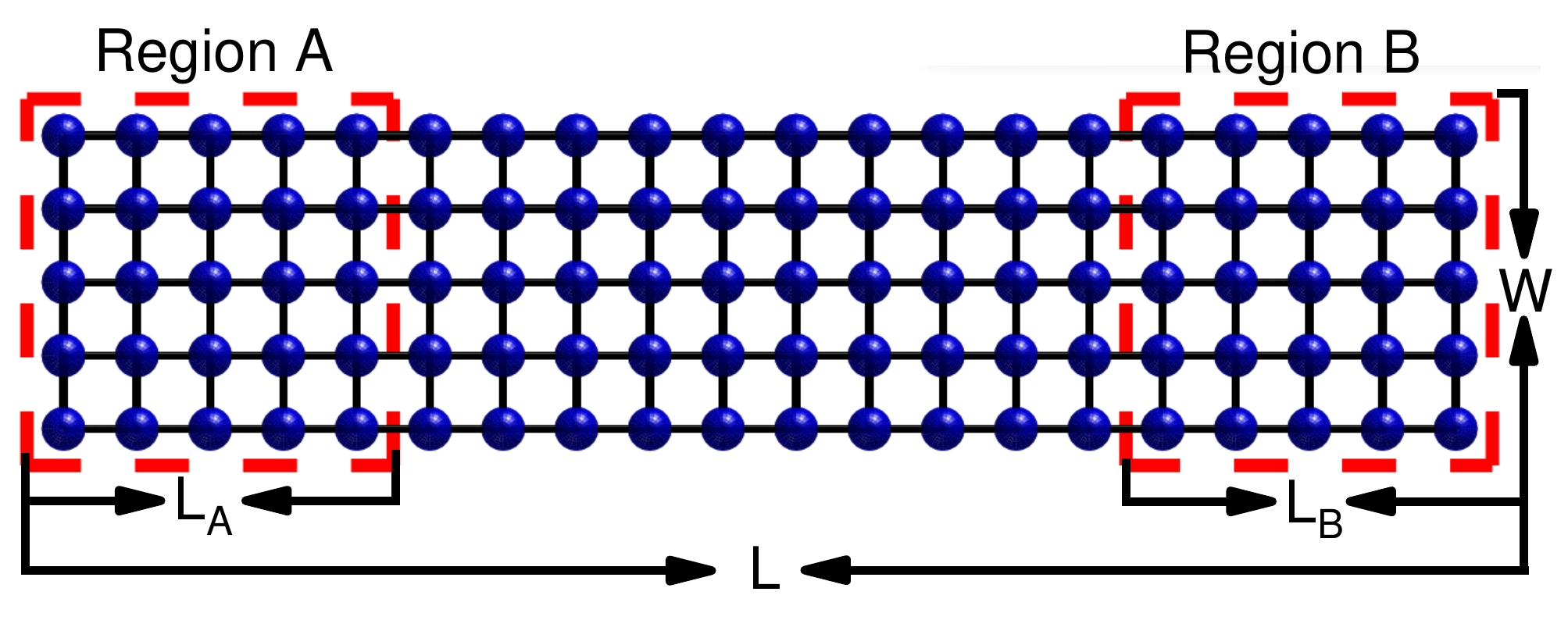}
	\caption{\textbf{Spatial partition for entanglement topological invariant computation}. Schematic diagram of the nanoflake partition with width $W$ and length $L$. The lengths of partition $A$ and $B$ are $L_{A}$ and $L_{B}$, respectively. And both of them  have the same width as the entire nanoflake.}
	\label{fig1}
\end{figure}
We construct the ETI of a finite size system as:
\begin{align}
  S^{T} = S_{A} + S_{B} - S_{\overline{A \cup B}},
  \label{eq1}
\end{align}
where $S_{A}$ and $S_{B}$ denote the entanglement entropies of two spatially separated boundary segments $A$ and $B$, respectively, and $S_{\overline{A \cup B}}$ represents the entanglement entropy of the complement of their union.
The expression of $S^T$ in Eq.~(\ref{eq1}) is constructed based on mutual information, but in free-fermion systems with higher-order topology, its role is especially dependent on the distance between regions A and B.
Unlike the conventional quantum mutual information, which measures the total correlations between two subsystems, typically defined on a closed manifold \cite{Wen2016}, the proposed ETI is specifically designed to isolate the topological contribution of entanglement in finite-size free-fermion systems under open boundaries. In particular, in free-fermion systems, the ground state is Gaussian and fully characterized by the two-point correlation functions. When selecting A and B as non-adjacent boundary segments (see Fig. \ref{fig1}) and maximizing their spatial separation, local correlations-whether from bulk or extended edge modes-decay rapidly and become negligible. Under this spatial separation condition, the remaining contribution to $S^{T}$ primarily reflects correlations between spatially localized corner states. Thus, $S^T$ can serve as a sensitive and robust indicator to characterize second-order topological insulators, where higher-order corner states contribute to the non-zero ETI. Consequently, the magnitude of $S^{T}$ correlates with the number of corner states and serves as a sensitive and robust indicator of higher-order topology. Therefore, while mathematically related to the quantum mutual information, the ETI represents a physically distinct quantity that captures the entanglement signature of higher-order topology in real-space systems.

Here, we would like to point out that the $S^{T}$ incorporates the core concepts of the quantum conditional mutual information (QCMI), effectively isolating long-range quantum correlations through a multi-partition spatial division \cite{Maiellaro2022,Maiellaro2022b,Maiellaro2023,Maiellaro2023b}. While QCMI measures total correlations across multiple partitions, it can be sensitive to system symmetries and may fail when zero-energy states are highly localized on a single side. In contrast, $S^{T}$ adopts the core idea of isolating long-range correlations but introduces a crucial modification: the entanglement entropy is computed using a maximally mixed ground-state ensemble. This approach effectively filters out short-range bulk correlations, reduces dependence on system symmetries, and reliably detects the presence and number of zero-dimensional topological states---even in cases with unilaterally localized edge modes. As a result, $S^{T}$ serves as a robust real-space invariant that can be broadly applied to free-fermion systems in one, two, and three dimensions, extending the applicability beyond the original scope of QCMI.
\\
\quad\\
\textbf{The application of  the ETI}\\
The bilayer-modified Bernevig-Hughes-Zhang (BHZ) model is employed to realize second-order topological insulators\cite{Liu2024}. This model serves as an ideal test case for verifying the effectiveness of ETI. The Hamiltonian consists of a top layer, $H_{T}(k)$, a bottom layer, $H_{B}(k)$, and interlayer coupling, $H_{C}$. Key features of the system include intra-orbital hopping, spin-orbit coupling, and an out-of-plane Zeeman field. This system supports corner-localized modes, which are characteristic of second-order topology, thus making it particularly suitable for investigating the entanglement properties of such phases. The coupled bilayer Hamiltonian in momentum space is expressed as follows \cite{Liu2024}:
\begin{align}
H(k) = \begin{pmatrix}
H_{T}(k) & H_{C} \\
H_{C}^{*} & H_{B}(k)
\end{pmatrix}.
\label{eq2}
\end{align}
The Hamiltonian components $H_{T}(k)$, $H_{B}(k)$, and $H_{C}$ can be written as:
\begin{align}
H_{T}(k) =& (t k_{x}^{2} + t k_{y}^{2} + \varepsilon) \sigma_{z} + \lambda_{x} k_{x} \sigma_{x} s_{z} + \lambda_{y} k_{y} \sigma_{y} s_{0}\nonumber \\
&+ B_{z} \sigma_{z} s_{z}, \nonumber \\
H_{B}(k) =& (t k_{x}^{2} + t k_{y}^{2} + \varepsilon) \sigma_{z} + \lambda_{x} k_{y} \sigma_{x} s_{z} + \lambda_{y} k_{x} \sigma_{y} s_{0}\nonumber \\
&+ B_{z} \sigma_{z} s_{z},\nonumber \\
H_{C} =&\eta\sigma_{0} s_{0},
\label{eq3}
\end{align}
where $t$ represents the nearest-neighbor intra-orbital hopping, $\varepsilon$ denotes the mass or gap parameter, and $\lambda_{x,y}$ indicates the kinetic energy. $ B_{z} $ and $ \eta $ correspond to the strengths of the out-of-plane Zeeman field and the interlayer coupling, respectively. Here,
$\sigma_{i}$ and $s_{i}$ for $i = ( x, y, z)$ are the Pauli matrices acting on the orbital and spin degrees of freedom, respectively, and $\sigma_{0}$ and $s_{0}$ represent the identity matrices for the orbital and spin spaces.

\begin{figure}
	\centering
	\includegraphics[width=1\columnwidth,clip]{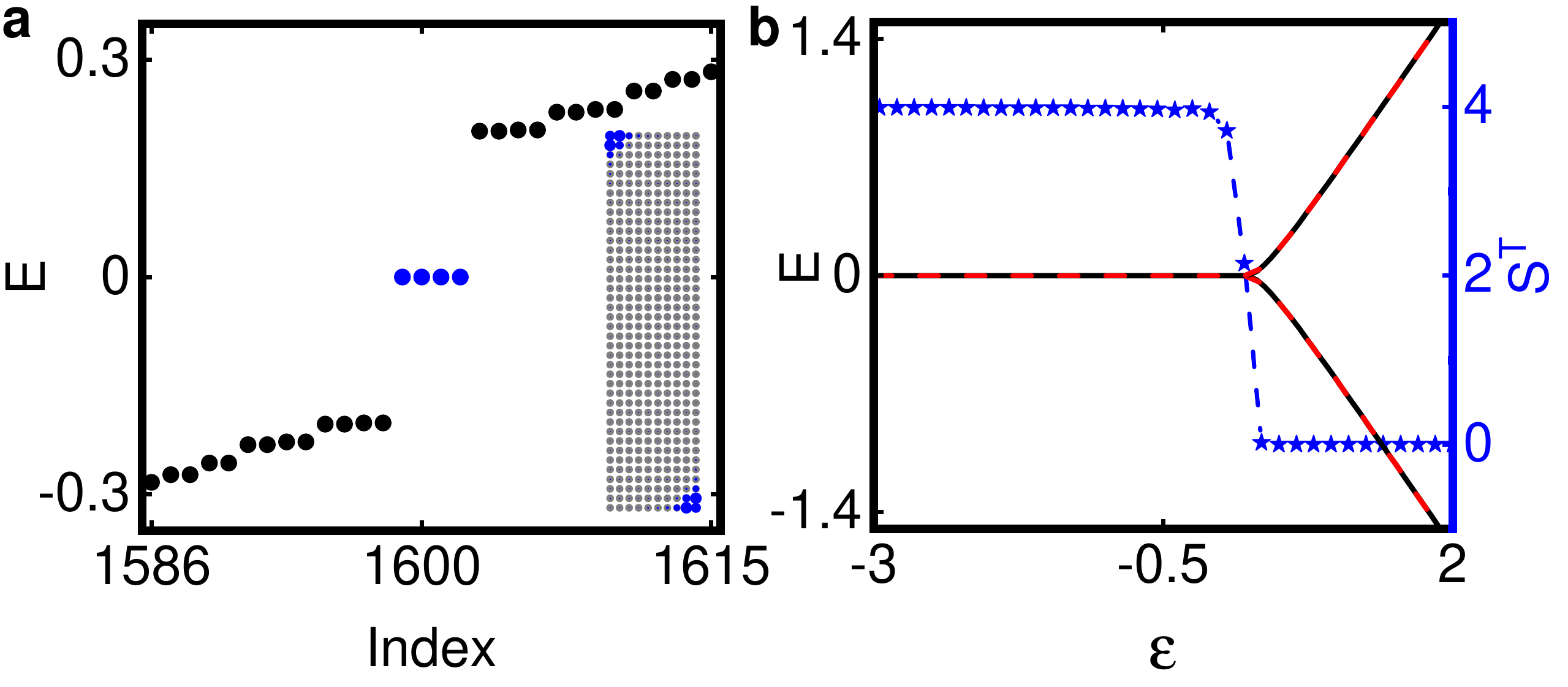}
	\caption{\textbf{Topological corner states and entanglement topological invariant evolution}. \textbf{a,} Energy levels of the square-shaped nanoflake for the coupled BHZ model with mass $\varepsilon=-1$.  The inset shows the probability density distribution of the zero-energy states represented by blue dots. \textbf{b,}  The energy $E$ near Fermi energy (black and red lines) and second-order ETI $S^{T}$ (blue star lines) as a function of mass $\varepsilon$. The alternating red and black lines represent the double degenerate states.
The other parameters are set as system width $W=10a$, system length $L=40a$, subsystem lengths $L_{A}=L_{B}=8a$, intra-orbital hopping $t=1$, kinetic energy $\lambda_{x}=\lambda_{y}=1$, interlayer coupling $\eta=0.4$ and Zeeman field $B_{z}=0$.
	}
	\label{fig2}
\end{figure}
To examine the behavior of $S^{T}$ during the phase transitions in the coupled BHZ model at finite system sizes, we perform numerical simulations with the following parameters. The nanoflake size is $L \times W=40a \times 10a$, where $a$ is the lattice constant.
The lengths of partitions $A$ and $B$ are set to $L_{A} = 8a$ and $L_{B} = 8a$, as illustrated in Fig. \ref{fig1}.
The remaining parameters are chosen as $B_{z}=0$, $\lambda_{x}=\lambda_{y}=1$, $\eta=0.4$, and $ t=1 $. In Fig. \ref{fig2}\textbf{a}, we show the energy spectrum of the nanoflake system for $\varepsilon =-1$. The system exhibits four zero-energy states (denoted by blue symbols) located within the bandgap of edge states. These states are primarily localized at the upper left and lower right corners of the sample, protected by time-reversal symmetry and mirror symmetry. This behavior demonstrates the characteristic features of second-order topological corner states.

As $\varepsilon$ varies within the range $[-3, 2]$, the fourfold zero-energy degeneracy at the Fermi level is lifted when $\varepsilon > 0$ (see the black and red lines in Fig. \ref{fig2}\textbf{b}), signaling a transition from a topological insulator into a trivial phase. Correspondingly, ETI, $S^{T}$, also experiences a sharp transition from $S^{T}=4$ in the topological phase to $S^{T}=0$ in the trivial phase (see the blue dashed line with star markers in Fig. \ref{fig2}\textbf{b}). This abrupt change occurs precisely at the critical point $\varepsilon = 0$, establishing a clear and direct correspondence between $S^{T}$ and the topological phase. Notably, once the system enters the topological phase, the value of $S^{T}$ precisely matches the number of zero-energy states. This indicates that the discrete values of the ETI likewise capture the count of zero-energy states, thereby providing a clear and intuitive indicator for identifying topological insulators and the emergence of zero-energy states. These findings demonstrate that $S^{T}$ effectively captures the correlation information of higher-order topological states and serves as a robust method for characterizing topological insulators, particularly in systems undergoing phase transitions.

The phase transition process described above does not involve any first-order topological phases. Although the ETI $S^{T}$ has some capability in characterizing first-order topological insulators, accurately capturing and describing such phases necessitates adjustments to the system's boundary conditions. Specifically, the open boundary conditions in the vertical direction must be replaced with periodic ones, while the horizontal direction retains open boundaries. However, as this work primarily focuses on the second-order topological phase and their associated phase transitions, a detailed analysis of first-order topological phases falls beyond the scope of this study. In the following sections, we will continue to explore the application of $S^{T}$ in two-dimensional second-order insulators.

\begin{figure}
	\centering
	\includegraphics[width=1\columnwidth,clip]{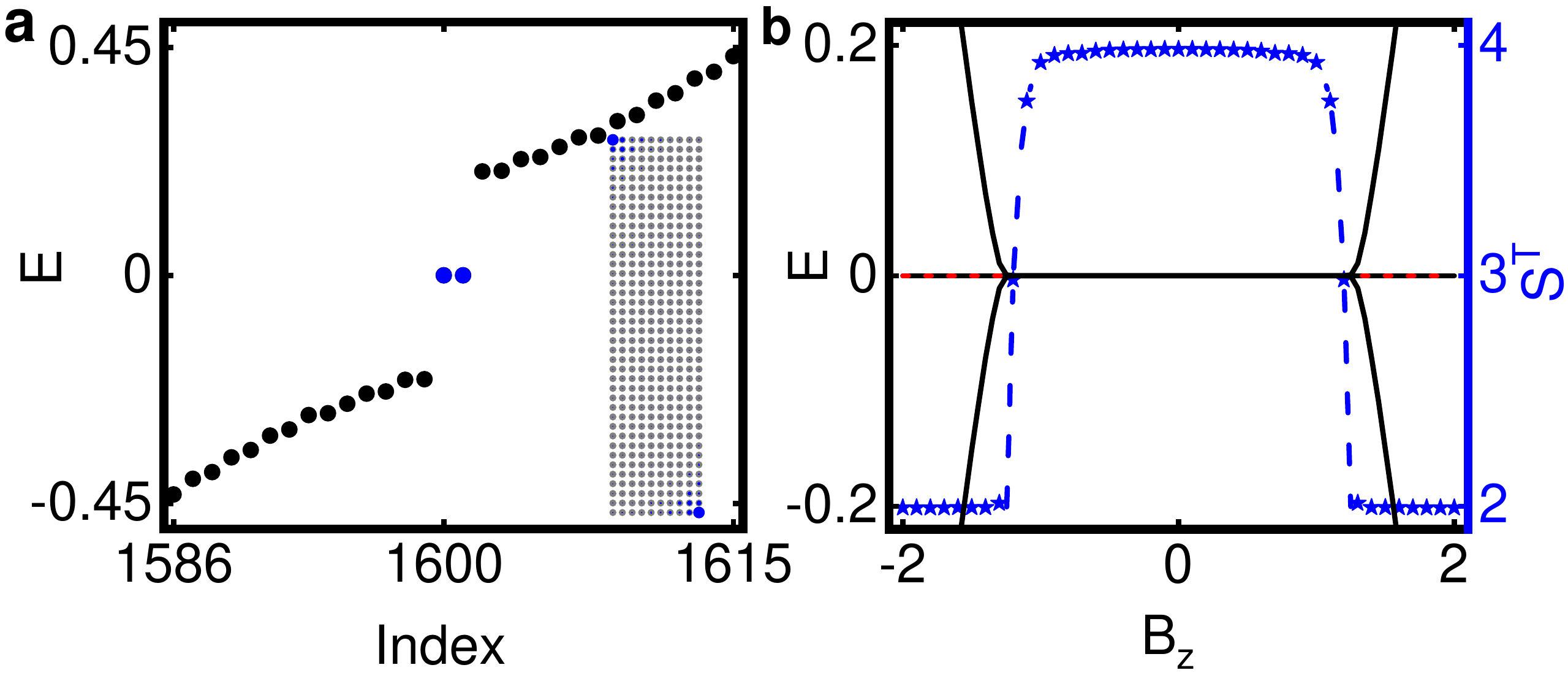}
	\caption{\textbf{Topological corner states and entanglement topological invariant evolution under exchange Zeeman field}. \textbf{a,}  Same as Fig. \ref{fig2}\textbf{a} , except the exchange Zeeman field $ B_{z} = 1.8 $. \textbf{b,}  The energy $E$ near Fermi energy (black and red lines) and second-order ETI $S^{T}$ (blue star lines) as a function of Zeeman field $B_{z}$. The alternating red and black lines represent the double degenerate states.
The other parameters are set as system width $W=10a$, system length $L=40a$, subsystem lengths $L_{A}=L_{B}=8a$, intra-orbital hopping $t=1$, kinetic energy $\lambda_{x}=\lambda_{y}=1$, interlayer coupling $\eta=0.4$ and mass $\varepsilon = -1$.
	}
	\label{fig3}
\end{figure}
To investigate the correspondence between the ETI value and the number of corner states, we introduce an out-of-plane Zeeman field into the coupled bilayer BHZ models. When $ B_{z} = 1.8 $, two zero-energy in-gap states emerge (see blue dots in Fig. \ref{fig3}\textbf{a}). Additionally, the probability distribution of the wave function at half-filling, shown in the inset of Fig. \ref{fig3}\textbf{a}, indicates that the two zero-energy in-gap states are primarily localized at the corners of the sample. By continuously tuning $B_{z}$ within the range $[-2,2]$, we observe that the number of zero-energy states (i.e., corner states) initially increases from 2 to 4, then decreases back to 2 (see the black and red lines in Fig. \ref{fig3}\textbf{b}). Correspondingly, the ETI $S^{T}$ transitions from 2 to 4, stabilizes at 4 when $B_{z}$ is near zero, and then returns to 2 (see the blue dashed line with star markers in Fig. \ref{fig3}\textbf{b}), reflecting a trend that mirrors the changes in the number of corner states. The information about the number of corner states can be well captured by the value of ETI.

\begin{figure}
	\centering
	\includegraphics[width=1\columnwidth,clip]{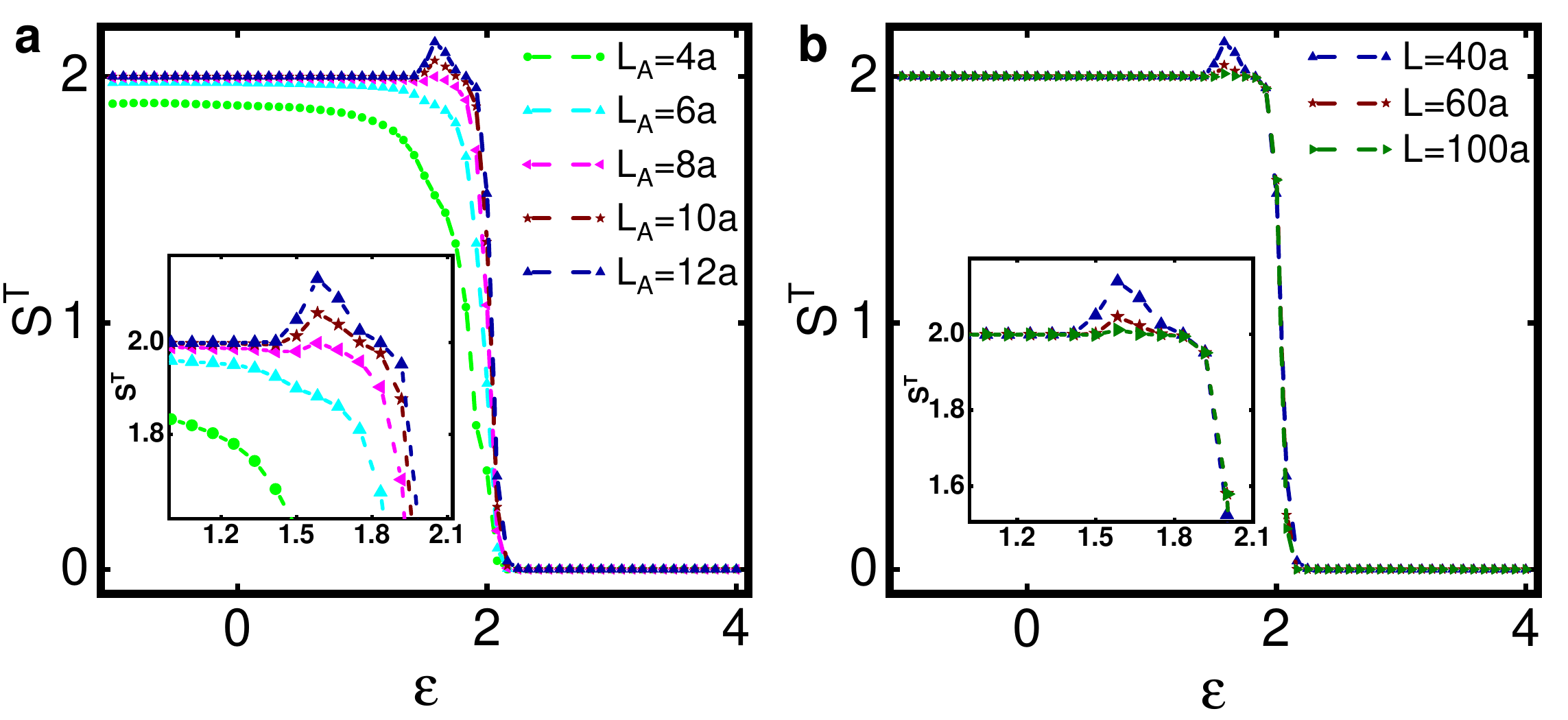}
	\caption{\textbf{Dependence of entanglement topological invariant on subsystem and system size}. Second-order ETI $S^{T}$ as a function of mass $\varepsilon$ with different subsystem lengths $L_{A}$ for \textbf{a}  and different system lengths $L$ for \textbf{b} . The length of region $B$ is consistently equal to the length of region $A$. The parameters are system length $L=40a$ in \textbf{a} and subsystem length $L_{A}=12a$ in \textbf{b} , other parameters are the same as those in Fig. \ref{fig2}\textbf{b}  except Zeeman field $ B_{z} = 1.8 $. The insets are the details of $ S^{T} $ within the range of mass $\varepsilon \in [1, 2.1]$.
	}
	\label{fig4}
\end{figure}

In Fig. \ref{fig4}, we examine the influence of system size on the ETI $S^{T}$ by analyzing its dependence on the parameter $\varepsilon$ for various subsystem sizes.
In Fig. \ref{fig4}\textbf{a}, we plot the ETI $S^{T}$ as function of $\varepsilon$ for various subsystem sizes $L_{A}$, where $L_{A}=L_{B}$, and the remaining parameters are consistent with those in Fig. \ref{fig2}\textbf{b}, except that $ B_{z}$ is set to $1.8 $. For $ L_{A} = 4a $, we observe that the $ S^{T} < 2$ in the nontrivial phase, while for larger subsystem sizes, the ETI $S^{T}$ stabilizes at a value of $2$. The reason is that the wave function of corner state is not strictly confined to a single point but remains localized within a finite region at the corner. If $ L_{A} $ is chosen too small, region $ A $ may fail to fully encompass the spatial extent of the corner-state wavefunction, potentially leading to an underestimation of its entanglement correlation.
However, for $ L_{A} = 10a $ and $ L_{A} = 12a $, near the phase transition point $\varepsilon = 2$, $ S^{T} $ exhibits significant fluctuations.
Specifically, for $ L_{A} = 12a $  the fluctuation amplitude is the largest, as shown in the inset graph in Fig. \ref{fig4}\textbf{a}.
To further investigate the fluctuations, in Fig. \ref{fig4}\textbf{b}, we plot the ETI $ S^{T} $ as a function of $\varepsilon$ for different system sizes ($L= 40a, 60a, 100a$) with fixed $ L_{A} =L_{B} = 12a $.
As observed in the inset graph in Fig. \ref{fig4}\textbf{b}), the peak before the phase transition disappears with the increasing system size. This phenomenon directly indicates that $ S^{T} $  is sensitive to short-range entanglement.
The above analysis suggests that the subsystem must be sufficiently large when using $ S^{T} $ as a second-order topological insulator indicator. Furthermore, once the subsystem size is chosen, maximizing the distance between subsystems is preferable. This can be expressed as $\lim_{L \to \infty} S^{T} = N_{0}$, where $N_{0}$ is the number of corner states.
Based on the research described above, we have successfully established a universal framework for characterizing the 2D second-order topological insulators in bilayer-modified BHZ model using the $S^{T}$ indicator. To further assess the applicability of this indicator in 2D second-order topological insulators, the Supplementary Information extends its application to six representative systems: (i) the two-dimensional Benalcazar-Bernevig-Hughes model (Supplementary Note 1), (ii) graphene nanoflakes with specific zigzag boundary conditions (Supplementary Note 2), (iii) the Kagome Model (Supplementary Note 3), (iv) the Kekul\'e lattice Model (Supplementary Note 4), (v) the one-dimensional Su-Schrieffer-Heeger model (Supplementary Note 5), and (vi) the three-dimensional Benalcazar-Bernevig-Hughes model (Supplementary Note 6).
Our calculations demonstrate that the $ S^{T} $ not only accurately identifies the critical point of the topological insulator but also numerically captures the degeneracy of the zero-energy bound states. This correspondence between $S^{T}$ and zero-energy bound states holds across various two-dimensional second-order topological systems, as well as in one-dimensional first-order systems and three-dimensional third-order systems, further confirming the broad applicability of our approach.
\\

\noindent
\textbf{\large Conclusions}\\
In this work, we propose a method for calculating the single-particle entanglement entropy in finite-sized two-dimensional systems. Building on this foundation, we introduce the ETI $S^{T}$, defined through bipartite entanglement, focusing on capturing corner ststes entanglement properties in such systems. By applying this framework to a coupled bilayer BHZ model, we establish a robust correspondence between $S^{T}$ and higher-order topological insulators.
Crucially, the value of $S^{T}$ quantitatively encodes the existence and degeneracy of corner-localized states, providing a direct, entanglement-based signature of higher-order topology. Importantly, this characterization holds even with the broken time-reversal symmetry, highlighting the versatility of $S^{T}$ across different symmetry regimes.
For appropriately chosen subsystem dimensions $L_{A}$, the asymptotic behavior $\lim_{L \to \infty} S^{T} = N_{0}$ emerges, where $N_{0}$ corresponds to the quantized count of two-dimensional corner states.
This result not only establishes $S^{T}$ as a complementary probe to conventional bulk topological invariants but also bridges the gap between entanglement-based diagnostics and real-space manifestations of higher-order topological insulators.

Meanwhile, in free-fermion systems, the invariant $S^{T}$ is computed directly from the single-particle correlation matrix and faithfully captures corner-localized zero modes. With weak interactions, the many-body ground state stays adiabatically linked to the non-interacting case if protecting symmetry and the bulk gap are maintained. Here, the invariant remains robust-topological boundary modes persist unless a phase transition occurs, and the entanglement structure still reflects their presence. In contrast, strong interactions can qualitatively change entanglement: they may gap out corner modes via spontaneous symmetry breaking or fractionalize them into emergent excitations like parafermions. In such regimes, the correlation-matrix approach fails, and a more general many-body treatment using the entanglement spectrum or symmetry-resolved entanglement entropy is required. Thus, while our entanglement invariant is well-defined for free fermions and stable under weak interactions, extending it to strongly correlated systems offers a promising direction for studying interacting higher-order topological phases.
\\

\noindent
\textbf{\large Methods}\\
\textbf{Entanglement entropy of finite-size free-fermion systems}\\
For a bipartite system $A \cup B$,  the total system's density matrix $\rho$ is generally expressed in the Gibbs form $\rho = \frac{e^{-\omega H}}{Z},$
where \(\omega = 1/T\) is a finite inverse temperature, \(H\) is the Hamiltonian of the system, and \(Z = \mathrm{Tr}(e^{-\omega H})\) is the partition function ensuring normalization. In the zero-temperature limit (\(T \to 0\)),
the exponential term $e^{-\omega H}$ in the Gibbs state converges to the ground state projector $\Pi_{gs}$:
\begin{align}
\rho \overset{\text{def}}{=} \frac{\Pi_{gs}}{\mathrm{Tr} \Pi_{gs}}.
  \label{eq4}
 \end{align}
 The entanglement entropy of subsystem $A$ is  then defined via the von Neumann entropy of the reduced density matrix, $\rho_{A} = \mathrm{Tr}_{B}(\rho)$, as:
\begin{align}
  S_{A} = -\text{Tr}(\rho_{A} \log \rho_{A}).
  \label{eq5}
 \end{align}

For non-interacting fermionic systems, the entanglement entropy $S_{A}$ can be efficiently computed using the eigenvalues of the correlation matrix
\cite{Peschel2009,Peschel2003}:
\begin{align}
  S_{A} = \sum_{i} -\zeta_{i} \ln \zeta_{i} - (1 - \zeta_{i}) \ln (1 - \zeta_{i}),
  \label{eq6}
 \end{align}
 where $\zeta_{i}$ denotes the $i$-th eigenvalue of correlation matrix ($C_{A}$) associated with subsystem $A$.

For the derivation of the correlation matrix, periodic boundary conditions are commonly employed due to their mathematical convenience. For a translationally invariant system, the Hamiltonian can be expressed in momentum space as:
\begin{align}
  H(k) = \sum_{\alpha, \beta} \hat{c}_{\alpha}^{\dagger} H_{\alpha, \beta}(k) \hat{c}_{\beta},
  \label{eq7}
 \end{align}
where $\hat{c}_{\alpha}^{\dagger}$($\hat{c}_{\beta}$) are the creation (annihilation) operators in the momentum representation, and $\alpha$, $\beta$ index the sublattice, orbital, spin, and other internal degrees of freedom. The correlation matrix in momentum space is given by
\begin{align}
  [C(k)]_{\alpha,\beta} = \langle \hat{c}_{\alpha}^{\dagger} \hat{c}_{\beta} \rangle = \sum_{n \in occ} \left[ u^{n}(k) \right]_{\alpha}^{*} \cdot \left[ u^{n}(k) \right]_{\beta} ,
  \label{eq8}
 \end{align}
where the summation runs over all occupied states (${occ}$), and $\left[ u^{n}(k) \right]_{\alpha}$ denotes the $\alpha$-th component of the $n$-th eigenvector of $H(k)$.  By performing a Fourier transform, the real-space correlation matrix $[C_{A}]_{r\alpha,r^{\prime}\beta}$ is obtained as
\begin{align}
  [C_{A}]_{r\alpha,r^{\prime}\beta} = \frac{1}{V_{\text{BZ}}} \int_{\text{BZ}} d^{d} k \, e^{-ik \cdot (r - r^{\prime})} [C(k)]_{\alpha,\beta},
  \label{eq9}
 \end{align}
where the integration is taken over the first Brillouin zone (BZ). To ensure locality, the indices $r$ and $r^{\prime}$ are restricted to the region $A$.

However, the implicit closed-loop geometry of the $A \cup B$ system in this approach fundamentally contradicts the finite size of actual physical systems. This discrepancy becomes particularly problematic when investigating topological insulators, as periodic boundary conditions may obscure the critical contributions from edge states. Consequently, the entanglement entropy obtained under these conditions may fail to fully capture the system's topological properties.

To address these limitations, it is essential to calculate the entanglement entropy in finite-size systems.
This requires constructing the correlation matrix directly in real space. The absence of translational symmetry renders momentum-space methods inapplicable, and the system's Hamiltonian and correlation matrix must be expressed as:
\begin{align}
&H = \sum_{r\alpha,r^{\prime}\beta} \hat{c}_{r\alpha}^{\dagger} H_{r\alpha,r^{\prime}\beta} \hat{c}_{r^{\prime}\beta},\nonumber \\
&[C_{A}]_{r\alpha,r^{\prime}\beta} = \langle \hat{c}_{r\alpha}^{\dagger} \hat{c}_{r^{\prime}\beta} \rangle.
\label{eq10}
\end{align}
where $\hat{c}_{r\alpha}^\dagger$ ($\hat{c}_{r'\beta}$) denote the creation (annihilation) operators in real space. When computing the correction matrix $C_A$, it is sufficient to restrict the indices
$r$ and $r'$ to the subsystem region $A$.
Importantly, the presence of edge states in finite-size systems leads to nontrivial modifications of the ground state, which significantly affects the evaluation of
$\langle \hat{c}_{r\alpha}^\dagger \hat{c}_{r'\beta} \rangle$. In the following, we provide a detailed derivation of the correlation matrix in finite-size systems and analyze its implications for entanglement entropy calculations.

The Hamiltonian can be diagonalized as:
\begin{align}
  H = \sum_{n} \hat{f}_{n}^{\dagger} E_{n} \hat{f}_{n},
  \label{eq11}
 \end{align}
where $ \hat{f}^\dagger_{n} = \sum_{r\alpha} \hat{c}_{r\alpha}^\dagger [u^{n}]_{r\alpha} $, and $[u^{n}]_{r\alpha}$ denotes the ${r\alpha}-$th component of the eigenvector of $H$ corresponding to eigenenergy $E_{n}$. The expectation value $\langle \hat{c}_{r\alpha}^{\dagger} \hat{c}_{r^{\prime}\beta} \rangle$ can be derived as:
 \begin{align}
  \langle \hat{c}_{r\alpha}^{\dagger} \hat{c}_{r^{\prime}\beta} \rangle
  &=\sum_{n,n^{\prime}} [u^{n}]_{r\alpha}^{*} \cdot [u^{n^{\prime}}]_{r^{\prime}\beta} \langle \hat{f}_n^{\dagger} \hat{f}_{n^{\prime}}  \rangle \nonumber \\
  &= \sum_{n} [u^{n}]_{r\alpha}^{*} \cdot [u^{n}]_{r^{\prime}\beta} \langle  \hat{f}_{n}^{\dagger} \hat{f}_{n}  \rangle.
  \label{eq12}
 \end{align}
Here,  $\langle  \hat{f}_{n}^{\dagger} \hat{f}_{n}  \rangle$ represents the expectation value of the occupation number for a state with eigenvalue
 $E_{n}$ in the ground state.
If the eigenenergies $ E_{n} $ of $ H $ are sorted in ascending order, then at half-filling, the first $\frac{N}{2}$ states are occupied, where
$N$ is the total number of states. Consequently, the expectation value simplifies to:
  \begin{align}
    \langle \hat{c}_{r\alpha}^{\dagger} \hat{c}_{r^{\prime}\beta} \rangle = \sum_{n=1}^{\frac{N}{2}} [u^{n}]_{r\alpha}^{*} \cdot [u^{n}]_{r^{\prime}\beta}.
  \label{eq13}
 \end{align}

However, under the finite-size effect, the emergence of edge states  inherently induces a highly degenerate ground state manifold.
In the zero-temperature limit, the degenerate ground states can be described by the maximally-mixed ground-state \cite{Arad2024}, expressed as
$\rho =\frac{1}{d} \sum_{i=1}^{d} |{GS}_i\rangle \langle {GS}_i|$,
where \(d\) denotes the degeneracy and $|{GS}_i\rangle$ is the $i$-th degenerate ground states. Assuming each ground state may carry an arbitrary global phase, i.e., $|{GS}_i\rangle \rightarrow e^{i\phi_i}|{GS}_i\rangle$, these phases cancel in the density matrix as $|{GS}_i\rangle \langle {GS}_i| \rightarrow e^{i\phi_i} |{GS}_i\rangle \langle {GS}_i| e^{-i\phi_i} = |{GS}_i\rangle \langle {GS}_i|$, and thus have no effect on the calculation. Consequently, such phases can be safely ignored.
Notably, the maximally-mixed ground-state is a Gaussian state, whose correlation matrix eigenvalues are related to the entanglement properties of the system.
 To account for this, we consider a system with four degenerate zero-energy states as an example, labeled as $E_{\xi} = 0$ for $\xi=\frac{N}{2}-1, \frac{N}{2},\frac{N}{2}+1$ and $\frac{N}{2}+2$, and modify Eq. (\ref{eq13}) accordingly.

The ground state exhibits a sixfold degeneracy, corresponding energetically equivalent occupation configurations, spanning an irreducible degenerate subspace in the Hilbert space:
 \begin{align}
    \{ &\hat{f}_{\frac{N}{2}-1}^{\dagger}\hat{f}_{\frac{N}{2}}^{\dagger}|g\rangle, \hat{f}_{\frac{N}{2}-1}^{\dagger}\hat{f}_{\frac{N}{2}+1}^{\dagger}|g\rangle,
     \hat{f}_{\frac{N}{2}-1}^{\dagger}\hat{f}_{\frac{N}{2}+2}^{\dagger}|g\rangle, \nonumber \\
    &\hat{f}_{\frac{N}{2}}^{\dagger}\hat{f}_{\frac{N}{2}+1}^{\dagger}|g\rangle,
     \hat{f}_{\frac{N}{2}}^{\dagger}\hat{f}_{\frac{N}{2}+2}^{\dagger}|g\rangle, \hat{f}_{\frac{N}{2}+1}^{\dagger}\hat{f}_{\frac{N}{2}+2}^{\dagger}|g\rangle
     \},
  \label{eq14}
 \end{align}
where $|g\rangle$  denoting  state with all negative - energy states occupied. The expectation value $\langle \hat{f}_{n}^{\dagger} \hat{f}_{n} \rangle$  is averaged over all six orthogonal ground states $\{|{GS}_i\rangle\}$ in this subspace:
\begin{equation}
\langle \hat{f}_{n}^{\dagger} \hat{f}_{n} \rangle = \frac{1}{6}\sum_{i=1}^6 \langle {GS}_i | \hat{f}_{n}^{\dagger} \hat{f}_{n} | {GS}_i \rangle.
  \label{eq15}
\end{equation}
For the entire system,   $\langle  \hat{f}_{n}^{\dagger} \hat{f}_{n}  \rangle$ can be expressed as:
\begin{align}
    \langle  \hat{f}_{n}^{\dagger} \hat{f}_{n}  \rangle = \left\{\begin{matrix}1, &n < \frac{N}{2}-1 \\
   \frac{1}{2}, &\frac{N}{2}-1 \leq n \leq \frac{N}{2}+2\\
   0, &n > \frac{N}{2}+2.\end{matrix}\right.
  \label{eq16}
 \end{align}
 Consequently, Eq. (\ref{eq13}) is modified as:
 \begin{align}
  [C_{A}]_{r\alpha,r^{\prime}\beta} = \sum_{n=1}^{\frac{N}{2}-2} [u^{n}]_{r\alpha}^{*} \cdot [u^{n}]_{r^{\prime}\beta} + \frac{1}{2} \sum_{n=\xi}[u^{n}]_{r\alpha}^{*} \cdot [u^{n}]_{r^{\prime}\beta},
  \label{eq17}
 \end{align}
where $\xi=\frac{N}{2}-1, \frac{N}{2},\frac{N}{2}+1,\frac{N}{2}+2$ denotes the indices of the four degenerate zero-energy states.
The above equation can be readily extended to systems with an arbitrary number of zero-energy degenerate states.
Subsequently, the bipartite entanglement entropy can be rigorously determined by substituting Eq. (\ref{eq17}) into Eq. (\ref{eq6}) for the finite size systems.
\quad\\	
\\
\textbf{\large Data availability}\\
The data that support the findings of this study are available from the corresponding author upon reasonable request. 
\quad\\
\\	
\textbf{\large Code availability}\\
The codes associated with this manuscript are available from the corresponding author upon reasonable request.
\quad\\
\\

\quad\\	
\\
\textbf{\large Acknowledgements}\\
This work was supported by the National Natural Science Foundation of China (Grants No. 12274305, No. 12074097, No. 12374034 and No. 12447147), Natural Science Foundation of Hebei Province (Grant No. A2024205025), the National Key R and D Program of China (Grant No. 2024YFA1409002), the China Postdoctoral Science Foundation (Grant No. 2024M760070), and
the Quantum Science and Technology-National Science and Technology Major Project (Grant No. 2021ZD0302403).\\

\quad\\	
\\
\textbf{\large Author contributions}\\
Y. -L. Z. and Y.-T. Z. conceived the work and designed the research strategy. Y.-L. Z. and C.-M. M. carried out the numerical calculations under the supervision of Y.-T. Z. and Q.-F. S.  All authors, Y.-L. Z., C.-M. M., J.-J. L., Y.-T. Z. and Q.-F. S. performed the data analysis and wrote the paper together.

\quad\\	
\\
\textbf{\large Competing interests}\\
The authors declare no competing interests.


\textbf{\large References}
\begin{thebibliography}{99}


        \expandafter\ifx\csname url\endcsname\relax
        \def\url#1{\texttt{#1}}\fi
        \expandafter\ifx\csname urlprefix\endcsname\relax\def\urlprefix{URL }\fi
        \providecommand{\bibinfo}[2]{#2}
        \providecommand{\eprint}[2][]{\url{#2}}



\bibitem{Song2017}
	\bibinfo{author}{Song, Z.}, \bibinfo{author}{Fang, Z.} \& \bibinfo{author}{Fang, C.}
	\newblock \bibinfo{title}{(d-2)-Dimensional Edge States of Rotation Symmetry Protected Topological States}.
	\newblock \emph{\bibinfo{journal}{Phys. Rev. Lett.}} \textbf{\bibinfo{volume}{119}}, \bibinfo{pages}{246402} (\bibinfo{year}{2017}).
	\newblock \urlprefix\url{https://journals.aps.org/prl/abstract/10.1103/PhysRevLett.119.246402}.
	
\bibitem{Langbehn2017}
	\bibinfo{author}{Langbehn, J.}, \bibinfo{author}{Peng, Y.}, \bibinfo{author}{Trifunovic, L.}, \bibinfo{author}{von Oppen, F.} \& \bibinfo{author}{Brouwer, P.~W.}
	\newblock \bibinfo{title}{Reflection-Symmetric Second-Order Topological Insulators and Superconductors}.
	\newblock \emph{\bibinfo{journal}{Phys. Rev. Lett.}} \textbf{\bibinfo{volume}{119}}, \bibinfo{pages}{246401} (\bibinfo{year}{2017}).
	\newblock \urlprefix\url{https://journals.aps.org/prl/abstract/10.1103/PhysRevLett.119.246401}.
	
\bibitem{Schindler2018}
	\bibinfo{author}{Schindler, F.} \emph{et~al.}
	\newblock \bibinfo{title}{Higher-order topological insulators}.
	\newblock \emph{\bibinfo{journal}{Sci. Adv.}} \textbf{\bibinfo{volume}{4}}, \bibinfo{pages}{eaat0346} (\bibinfo{year}{2018}).
	\newblock \urlprefix\url{https://www.science.org/doi/10.1126/sciadv.aat0346}.
	
\bibitem{Franca2018}
	\bibinfo{author}{Franca, S.}, \bibinfo{author}{van den Brink, J.} \& \bibinfo{author}{Fulga, I.~C.}
	\newblock \bibinfo{title}{An anomalous higher-order topological insulator}.
	\newblock \emph{\bibinfo{journal}{Phys. Rev. B}} \textbf{\bibinfo{volume}{98}}, \bibinfo{pages}{201114} (\bibinfo{year}{2018}).
	\newblock \urlprefix\url{https://journals.aps.org/prb/abstract/10.1103/PhysRevB.98.201114}.
	
\bibitem{Wang2019}
	\bibinfo{author}{Wang, Z.} \emph{et~al.}
	\newblock \bibinfo{title}{Higher-Order Topology, Monopole Nodal Lines, and the Origin of Large Fermi Arcs in Transition Metal Dichalcogenides XTe$_2$ (X = Mo, W)}.
	\newblock \emph{\bibinfo{journal}{Phys. Rev. Lett.}} \textbf{\bibinfo{volume}{123}}, \bibinfo{pages}{186401} (\bibinfo{year}{2019}).
	\newblock \urlprefix\url{https://journals.aps.org/prl/abstract/10.1103/PhysRevLett.123.186401}.
	
\bibitem{Ezawa2018}
	\bibinfo{author}{Ezawa, M.}
	\newblock \bibinfo{title}{Higher-Order Topological Insulators and Semimetals on the Breathing Kagome and Pyrochlore Lattices}.
	\newblock \emph{\bibinfo{journal}{Phys. Rev. Lett.}} \textbf{\bibinfo{volume}{120}}, \bibinfo{pages}{026801} (\bibinfo{year}{2018}).
	\newblock \urlprefix\url{https://journals.aps.org/prl/abstract/10.1103/PhysRevLett.120.026801}.
	
\bibitem{Calugaru2019}
	\bibinfo{author}{C\u{a}lug\u{a}ru, D.}, \bibinfo{author}{Juri\u{c}i\'{c}, V.} \& \bibinfo{author}{Roy, B.}
	\newblock \bibinfo{title}{Higher-order topological phases: A general principle of construction}.
	\newblock \emph{\bibinfo{journal}{Phys. Rev. B}} \textbf{\bibinfo{volume}{99}}, \bibinfo{pages}{041301} (\bibinfo{year}{2019}).
	\newblock \urlprefix\url{https://journals.aps.org/prb/abstract/10.1103/PhysRevB.99.041301}.
	
\bibitem{Trifunovic2019}
	\bibinfo{author}{Trifunovic, L.} \& \bibinfo{author}{Brouwer, P.~W.}
	\newblock \bibinfo{title}{Higher-Order Bulk-Boundary Correspondence for Topological Crystalline Phases}.
	\newblock \emph{\bibinfo{journal}{Phys. Rev. X}} \textbf{\bibinfo{volume}{9}}, \bibinfo{pages}{011012} (\bibinfo{year}{2019}).
	\newblock \urlprefix\url{https://journals.aps.org/prx/abstract/10.1103/PhysRevX.9.011012}.
	
\bibitem{Khalaf2018}
	\bibinfo{author}{Khalaf, E.}
	\newblock \bibinfo{title}{Higher-order topological insulators and superconductors protected by inversion symmetry}.
	\newblock \emph{\bibinfo{journal}{Phys. Rev. B}} \textbf{\bibinfo{volume}{97}}, \bibinfo{pages}{205136} (\bibinfo{year}{2018}).
	\newblock \urlprefix\url{https://journals.aps.org/prb/abstract/10.1103/PhysRevB.97.205136}.
	
\bibitem{addref1}
	\bibinfo{author}{Zeng, J.}, \bibinfo{author}{Liu, H.}, \bibinfo{author}{Jiang, H.}, \bibinfo{author}{Sun, Q.-F.} \& \bibinfo{author}{Xie, X.~C.}
	\newblock \bibinfo{title}{Multiorbital model reveals a second-order topological insulator in 1H transition metal dichalcogenides}.
	\newblock \emph{\bibinfo{journal}{Phys. Rev. B}} \textbf{\bibinfo{volume}{104}}, \bibinfo{pages}{L161108} (\bibinfo{year}{2021}).
	\newblock \urlprefix\url{https://journals.aps.org/prb/abstract/10.1103/PhysRevB.104.L161108}.

\bibitem{Wang2025}
    \bibinfo{author}{Wang, Z.} \emph{et~al.}
    \newblock \bibinfo{title}{Realization of a three-dimensional photonic higher-order topological insulator}.
    \newblock \emph{\bibinfo{journal}{Nat. Commun.}} \textbf{\bibinfo{volume}{16}}, \bibinfo{pages}{3122} (\bibinfo{year}{2025}).
    \newblock \urlprefix\url{https://www.nature.com/articles/s41467-025-58051-7}
	
\bibitem{Thouless1982}
	\bibinfo{author}{Thouless, D.~J.}, \bibinfo{author}{Kohmoto, M.}, \bibinfo{author}{Nightingale, M.~P.} \& \bibinfo{author}{den Nijs, M.}
	\newblock \bibinfo{title}{Quantized Hall Conductance in a Two-Dimensional Periodic Potential}.
	\newblock \emph{\bibinfo{journal}{Phys. Rev. Lett.}} \textbf{\bibinfo{volume}{49}}, \bibinfo{pages}{405--408} (\bibinfo{year}{1982}).
	\newblock \urlprefix\url{https://journals.aps.org/prl/abstract/10.1103/PhysRevLett.49.405}.
	
\bibitem{KaneMele2005a}
	\bibinfo{author}{Kane, C.~L.} \& \bibinfo{author}{Mele, E.~J.}
	\newblock \bibinfo{title}{Quantum Spin Hall Effect in Graphene}.
	\newblock \emph{\bibinfo{journal}{Phys. Rev. Lett.}} \textbf{\bibinfo{volume}{95}}, \bibinfo{pages}{226801} (\bibinfo{year}{2005}).
	\newblock \urlprefix\url{https://journals.aps.org/prl/abstract/10.1103/PhysRevLett.95.226801}.
	
\bibitem{KaneMele2005b}
	\bibinfo{author}{Kane, C.~L.} \& \bibinfo{author}{Mele, E.~J.}
	\newblock \bibinfo{title}{$\mathbb{Z}_2$ Topological Order and the Quantum Spin Hall Effect}.
	\newblock \emph{\bibinfo{journal}{Phys. Rev. Lett.}} \textbf{\bibinfo{volume}{95}}, \bibinfo{pages}{146802} (\bibinfo{year}{2005}).
	\newblock \urlprefix\url{https://journals.aps.org/prl/abstract/10.1103/PhysRevLett.95.146802}.
	
\bibitem{Benalcazar2017}
	\bibinfo{author}{Benalcazar, W.~A.}, \bibinfo{author}{Bernevig, B.~A.} \& \bibinfo{author}{Hughes, T.~L.}
	\newblock \bibinfo{title}{Quantized electric multipole insulators}.
	\newblock \emph{\bibinfo{journal}{Science}} \textbf{\bibinfo{volume}{357}}, \bibinfo{pages}{61} (\bibinfo{year}{2017}).
	\newblock \urlprefix\url{https://www.science.org/doi/10.1126/science.aah6442}.
	
\bibitem{Benalcazar2017PRB}
	\bibinfo{author}{Benalcazar, W.~A.}, \bibinfo{author}{Bernevig, B.~A.} \& \bibinfo{author}{Hughes, T.~L.}
	\newblock \bibinfo{title}{Electric multipole moments, topological multipole moment pumping, and chiral hinge states in crystalline insulators}.
	\newblock \emph{\bibinfo{journal}{Phys. Rev. B}} \textbf{\bibinfo{volume}{96}}, \bibinfo{pages}{245115} (\bibinfo{year}{2017}).
	\newblock \urlprefix\url{https://journals.aps.org/prb/abstract/10.1103/PhysRevB.96.245115}.
	
\bibitem{Ono2019}
	\bibinfo{author}{Ono, S.} \emph{et~al.}
	\newblock \bibinfo{title}{Difficulties in operator-based formulation of the bulk quadrupole moment}.
	\newblock \emph{\bibinfo{journal}{Phys. Rev. B}} \textbf{\bibinfo{volume}{100}}, \bibinfo{pages}{245133} (\bibinfo{year}{2019}).
	\newblock \urlprefix\url{https://journals.aps.org/prb/abstract/10.1103/PhysRevB.100.245133}.
	
\bibitem{Li2020}
	\bibinfo{author}{Li, C.-A.}\emph{et~al.}
	\newblock \bibinfo{title}{Topological insulators in disordered electric quadrupole insulators}.
	\newblock \emph{\bibinfo{journal}{Phys. Rev. Lett.}} \textbf{\bibinfo{volume}{125}}, \bibinfo{pages}{166801} (\bibinfo{year}{2020}).
	\newblock \urlprefix\url{https://doi.org/10.1103/PhysRevLett.125.166801}.
	
\bibitem{Schindler2018a}
	\bibinfo{author}{Schindler, F.} \emph{et~al.}
	\newblock \bibinfo{title}{Higher-order topology in bismuth}.
	\newblock \emph{\bibinfo{journal}{Nat. Phys.}} \textbf{\bibinfo{volume}{14}}, \bibinfo{pages}{918} (\bibinfo{year}{2018}).
	\newblock \urlprefix\url{https://www.nature.com/articles/s41567-018-0224-7}.
	
\bibitem{Costa2021}
	\bibinfo{author}{Costa, M.} \emph{et~al.}
	\newblock \bibinfo{title}{Discovery of higher-order topological insulators using the spin Hall conductivity as a topology signature}.
	\newblock \emph{\bibinfo{journal}{npj Comput. Mater.}} \textbf{\bibinfo{volume}{7}}, \bibinfo{pages}{146} (\bibinfo{year}{2021}).
	\newblock \urlprefix\url{https://www.nature.com/articles/s41524-021-00518-4}.
	
\bibitem{Miao2022}
	\bibinfo{author}{Miao, C.-M.}, \bibinfo{author}{Sun, Q.-F.} \& \bibinfo{author}{Zhang, Y.-T.}
	\newblock \bibinfo{title}{Second-order topological corner states in zigzag graphene nanoflake with different types of edge magnetic configurations}.
	\newblock \emph{\bibinfo{journal}{Phys. Rev. B}} \textbf{\bibinfo{volume}{106}}, \bibinfo{pages}{165422} (\bibinfo{year}{2022}).
	\newblock \urlprefix\url{https://journals.aps.org/prb/abstract/10.1103/PhysRevB.106.165422}.
	
\bibitem{Miao2023}
	\bibinfo{author}{Miao, C.-M.}, \bibinfo{author}{Wan, Y.-H.}, \bibinfo{author}{Sun, Q.-F.} \& \bibinfo{author}{Zhang, Y.-T.}
	\newblock \bibinfo{title}{Engineering topologically protected zero-dimensional interface end states in antiferromagnetic heterojunction graphene nanoflakes}.
	\newblock \emph{\bibinfo{journal}{Phys. Rev. B}} \textbf{\bibinfo{volume}{108}}, \bibinfo{pages}{075401} (\bibinfo{year}{2023}).
	\newblock \urlprefix\url{https://journals.aps.org/prb/abstract/10.1103/PhysRevB.108.075401}.
	
\bibitem{Miao2024}
	\bibinfo{author}{Miao, C.-M.}, \bibinfo{author}{Liu, L.}, \bibinfo{author}{Wan, Y.-H.}, \bibinfo{author}{Sun, Q.-F.} \& \bibinfo{author}{Zhang, Y.-T.}
	\newblock \bibinfo{title}{General principle behind magnetization-induced second-order topological corner states in the Kane-Mele model}.
	\newblock \emph{\bibinfo{journal}{Phys. Rev. B}} \textbf{\bibinfo{volume}{109}}, \bibinfo{pages}{205417} (\bibinfo{year}{2024}).
	\newblock \urlprefix\url{https://journals.aps.org/prb/abstract/10.1103/PhysRevB.109.205417}.
	
\bibitem{Yang2020}
	\bibinfo{author}{Yang, Y.-B.} \emph{et~al.}
	\newblock \bibinfo{title}{Type-II quadrupole topological insulators}.
	\newblock \emph{\bibinfo{journal}{Phys. Rev. Research}} \textbf{\bibinfo{volume}{2}}, \bibinfo{pages}{033029} (\bibinfo{year}{2020}).
	\newblock \urlprefix\url{https://journals.aps.org/prresearch/abstract/10.1103/PhysRevResearch.2.033029}.
	
\bibitem{Kitaev2006}
	\bibinfo{author}{Kitaev, A.} \& \bibinfo{author}{Preskill, J.}
	\newblock \bibinfo{title}{Topological Entanglement Entropy}.
	\newblock \emph{\bibinfo{journal}{Phys. Rev. Lett.}} \textbf{\bibinfo{volume}{96}}, \bibinfo{pages}{110404} (\bibinfo{year}{2006}).
	\newblock \urlprefix\url{https://journals.aps.org/prl/abstract/10.1103/PhysRevLett.96.110404}.
	
\bibitem{Levin2006}
	\bibinfo{author}{Levin, M.} \& \bibinfo{author}{Wen, X.-G.}
	\newblock \bibinfo{title}{Detecting Topological Order in a Ground State Wave Function}.
	\newblock \emph{\bibinfo{journal}{Phys. Rev. Lett.}} \textbf{\bibinfo{volume}{96}}, \bibinfo{pages}{110405} (\bibinfo{year}{2006}).
	\newblock \urlprefix\url{https://journals.aps.org/prl/abstract/10.1103/PhysRevLett.96.110405}.
	
\bibitem{Amico2008}
	\bibinfo{author}{Amico, L.}, \bibinfo{author}{Fazio, R.}, \bibinfo{author}{Osterloh, A.} \& \bibinfo{author}{Vedral, V.}
	\newblock \bibinfo{title}{Entanglement in many-body systems}.
	\newblock \emph{\bibinfo{journal}{Rev. Mod. Phys.}} \textbf{\bibinfo{volume}{80}}, \bibinfo{pages}{517} (\bibinfo{year}{2008}).
	\newblock \urlprefix\url{https://journals.aps.org/rmp/abstract/10.1103/RevModPhys.80.517}.
	
\bibitem{Calabrese2009}
	\bibinfo{author}{Calabrese, P.} \& \bibinfo{author}{Cardy, J.}
	\newblock \bibinfo{title}{Entanglement entropy and conformal field theory}.
	\newblock \emph{\bibinfo{journal}{J. Phys. A: Math. Theor.}} \textbf{\bibinfo{volume}{42}}, \bibinfo{pages}{504005} (\bibinfo{year}{2009}).
	\newblock \urlprefix\url{https://iopscience.iop.org/article/10.1088/1751-8113/42/50/504005}.
	
\bibitem{Eisert2010}
	\bibinfo{author}{Eisert, J.}, \bibinfo{author}{Cramer, M.} \& \bibinfo{author}{Plenio, M.~B.}
	\newblock \bibinfo{title}{Colloquium: Area laws for the entanglement entropy}.
	\newblock \emph{\bibinfo{journal}{Rev. Mod. Phys.}} \textbf{\bibinfo{volume}{82}}, \bibinfo{pages}{277} (\bibinfo{year}{2010}).
	\newblock \urlprefix\url{https://journals.aps.org/rmp/abstract/10.1103/RevModPhys.82.277}.
	
\bibitem{Fradkin2013}
	\bibinfo{author}{Fradkin, E.}
	\newblock \emph{\bibinfo{title}{Field Theories of Condensed Matter Physics}}.
	\newblock \bibinfo{publisher}{Cambridge University Press}, \bibinfo{address}{Cambridge} (\bibinfo{year}{2013}).
	\newblock \urlprefix\url{https://doi.org/10.1017/CBO9781139015509}.
	
\bibitem{Tsomokos2009}
	\bibinfo{author}{Tsomokos, D.~I.} \emph{et~al.}
	\newblock \bibinfo{title}{Topological order following a quantum quench}.
	\newblock \emph{\bibinfo{journal}{Phys. Rev. A}} \textbf{\bibinfo{volume}{80}}, \bibinfo{pages}{060302} (\bibinfo{year}{2009}).
	\newblock \urlprefix\url{https://journals.aps.org/pra/abstract/10.1103/PhysRevA.80.060302}.
	
\bibitem{Halasz2012}
	\bibinfo{author}{Hal\'{a}sz, G.~B.} \& \bibinfo{author}{Hamma, A.}
	\newblock \bibinfo{title}{Probing topological order with R\'{e}nyi entropy}.
	\newblock \emph{\bibinfo{journal}{Phys. Rev. A}} \textbf{\bibinfo{volume}{86}}, \bibinfo{pages}{062330} (\bibinfo{year}{2012}).
	\newblock \urlprefix\url{https://journals.aps.org/pra/abstract/10.1103/PhysRevA.86.062330}.
	
\bibitem{Halasz2013}
	\bibinfo{author}{Hal\'{a}sz, G.~B.} \& \bibinfo{author}{Hamma, A.}
	\newblock \bibinfo{title}{Topological R\'{e}nyi Entropy after a Quantum Quench}.
	\newblock \emph{\bibinfo{journal}{Phys. Rev. Lett.}} \textbf{\bibinfo{volume}{110}}, \bibinfo{pages}{170605} (\bibinfo{year}{2013}).
	\newblock \urlprefix\url{https://journals.aps.org/prl/abstract/10.1103/PhysRevLett.110.170605}.
	
\bibitem{Depenbrock2012}
	\bibinfo{author}{Depenbrock, S.}, \bibinfo{author}{McCulloch, I.~P.} \& \bibinfo{author}{Schollw\"{o}ck, U.}
	\newblock \bibinfo{title}{Nature of the Spin-Liquid Ground State of the $S$=1/2 Heisenberg Model on the Kagome Lattice}.
	\newblock \emph{\bibinfo{journal}{Phys. Rev. Lett.}} \textbf{\bibinfo{volume}{109}}, \bibinfo{pages}{067201} (\bibinfo{year}{2012}).
	\newblock \urlprefix\url{https://journals.aps.org/prl/abstract/10.1103/PhysRevLett.109.067201}.
	
\bibitem{Jiang2012}
	\bibinfo{author}{Jiang, H.-C.}, \bibinfo{author}{Wang, Z.} \& \bibinfo{author}{Balents, L.}
	\newblock \bibinfo{title}{Identifying topological order by entanglement entropy}.
	\newblock \emph{\bibinfo{journal}{Nat. Phys.}} \textbf{\bibinfo{volume}{8}}, \bibinfo{pages}{902} (\bibinfo{year}{2012}).
	\newblock \urlprefix\url{https://www.nature.com/articles/nphys2465}.
	
\bibitem{Isakov2011}
	\bibinfo{author}{Isakov, S.~V.}, \bibinfo{author}{Hastings, M.~B.} \& \bibinfo{author}{Melko, R.~G.}
	\newblock \bibinfo{title}{Topological Entanglement Entropy of a Bose-Hubbard Spin Liquid}.
	\newblock \emph{\bibinfo{journal}{Nat. Phys.}} \textbf{\bibinfo{volume}{7}}, \bibinfo{pages}{772} (\bibinfo{year}{2011}).
	\newblock \urlprefix\url{https://www.nature.com/articles/nphys2036}.
	
\bibitem{Sankar2023}
	\bibinfo{author}{Sankar, S.}, \bibinfo{author}{Sela, E.} \& \bibinfo{author}{Han, C.}
	\newblock \bibinfo{title}{Measuring Topological Entanglement Entropy Using Maxwell Relations}.
	\newblock \emph{\bibinfo{journal}{Phys. Rev. Lett.}} \textbf{\bibinfo{volume}{131}}, \bibinfo{pages}{016601} (\bibinfo{year}{2023}).
	\newblock \urlprefix\url{https://link.aps.org/doi/10.1103/PhysRevLett.131.016601}.

\bibitem{Karamlou2024}
	\bibinfo{author}{Karamlou, A. H.} \emph{et~al.}
	\newblock \bibinfo{title}{Probing entanglement in a 2D hard-core Bose--Hubbard lattice}.
	\newblock \emph{\bibinfo{journal}{Nature}} \textbf{\bibinfo{volume}{629}}, \bibinfo{pages}{561--566} (\bibinfo{year}{2024}).
	\newblock \url{https://www.nature.com/articles/s41586-024-07325-z}
	
\bibitem{Lin2024}
	\bibinfo{author}{Lin, Z.-K.} \emph{et~al.}
	\newblock \bibinfo{title}{Measuring entanglement entropy and its topological signature for phononic systems}.
	\newblock \emph{\bibinfo{journal}{Nat. Commun.}} \textbf{\bibinfo{volume}{15}}, \bibinfo{pages}{1601} (\bibinfo{year}{2024}).
	\newblock \urlprefix\url{https://www.nature.com/articles/s41467-024-45887-8}.

\bibitem{Chen2019}
    \bibinfo{author}{Chen, T.} \emph{et~al.}
    \newblock \bibinfo{title}{Experimental observation of classical analogy of topological entanglement entropy}.
    \newblock \emph{\bibinfo{journal}{Nat. Commun.}} \textbf{\bibinfo{volume}{10}}, \bibinfo{pages}{1557} (\bibinfo{year}{2019}).
    \newblock \urlprefix\url{https://www.nature.com/articles/s41467-019-09584-1}
	
\bibitem{Li2008}
	\bibinfo{author}{Li, H.} \& \bibinfo{author}{Haldane, F.~D.~M.}
	\newblock \bibinfo{title}{Entanglement Spectrum as a Generalization of Entanglement Entropy: Identification of Topological Order in Non-Abelian Fractional Quantum Hall Effect States}.
	\newblock \emph{\bibinfo{journal}{Phys. Rev. Lett.}} \textbf{\bibinfo{volume}{101}}, \bibinfo{pages}{010504} (\bibinfo{year}{2008}).
	\newblock \urlprefix\url{https://journals.aps.org/prl/abstract/10.1103/PhysRevLett.101.010504}.
	
\bibitem{Chandran2011}
	\bibinfo{author}{Chandran, A.} \emph{et~al.}
	\newblock \bibinfo{title}{Bulk-edge correspondence in entanglement spectra}.
	\newblock \emph{\bibinfo{journal}{Phys. Rev. B}} \textbf{\bibinfo{volume}{84}}, \bibinfo{pages}{205136} (\bibinfo{year}{2011}).
	\newblock \urlprefix\url{https://journals.aps.org/prb/abstract/10.1103/PhysRevB.84.205136}.
	
\bibitem{Prodan2010}
	\bibinfo{author}{Prodan, E.}, \bibinfo{author}{Hughes, T.~L.} \& \bibinfo{author}{Bernevig, B.~A.}
	\newblock \bibinfo{title}{Entanglement Spectrum of a Disordered Topological Chern Insulator}.
	\newblock \emph{\bibinfo{journal}{Phys. Rev. Lett.}} \textbf{\bibinfo{volume}{105}}, \bibinfo{pages}{115501} (\bibinfo{year}{2010}).
	\newblock \urlprefix\url{https://journals.aps.org/prl/abstract/10.1103/PhysRevLett.105.115501}.

\bibitem{Mao2025}
    \bibinfo{author}{Mao, B.-B.} \emph{et~al.}
    \newblock \bibinfo{title}{Sampling reduced density matrix to extract fine levels of entanglement spectrum and restore entanglement Hamiltonian}.
    \newblock \emph{\bibinfo{journal}{Nat. Commun.}} \textbf{\bibinfo{volume}{16}}, \bibinfo{pages}{2880} (\bibinfo{year}{2025}).
    \newblock \urlprefix\url{https://www.nature.com/articles/s41467-025-58058-0}


\bibitem{Wen2016}
     \bibinfo{author}{Wen, X.} \emph{et~al.}
     \newblock \bibinfo{title}{Edge theory approach to topological entanglement entropy, mutual information, and entanglement negativity in Chern-Simons       theories}.
     \newblock \emph{\bibinfo{journal}{Phys. Rev. B}} \textbf{\bibinfo{volume}{93}}, \bibinfo{pages}{245140} (\bibinfo{year}{2016}).
     \newblock \urlprefix\url{https://doi.org/10.1103/PhysRevB.93.245140}
 	
\bibitem{Fidkowski2010}
	\bibinfo{author}{Fidkowski, L.}
	\newblock \bibinfo{title}{Entanglement Spectrum of Topological Insulators and Superconductors}.
	\newblock \emph{\bibinfo{journal}{Phys. Rev. Lett.}} \textbf{\bibinfo{volume}{104}}, \bibinfo{pages}{130502} (\bibinfo{year}{2010}).
	\newblock \urlprefix\url{https://journals.aps.org/prl/abstract/10.1103/PhysRevLett.104.130502}.
	
\bibitem{Alexandradinata2011}
	\bibinfo{author}{Alexandradinata, A.}, \bibinfo{author}{Hughes, T.~L.} \& \bibinfo{author}{Bernevig, B.~A.}
	\newblock \bibinfo{title}{Trace index and spectral flow in the entanglement spectrum of topological insulators}.
	\newblock \emph{\bibinfo{journal}{Phys. Rev. B}} \textbf{\bibinfo{volume}{84}}, \bibinfo{pages}{195103} (\bibinfo{year}{2011}).
	\newblock \urlprefix\url{https://journals.aps.org/prb/abstract/10.1103/PhysRevB.84.195103}.
	
\bibitem{Hughes2011}
	\bibinfo{author}{Hughes, T.~L.}, \bibinfo{author}{Prodan, E.} \& \bibinfo{author}{Bernevig, B.~A.}
	\newblock \bibinfo{title}{Inversion-symmetric topological insulators}.
	\newblock \emph{\bibinfo{journal}{Phys. Rev. B}} \textbf{\bibinfo{volume}{83}}, \bibinfo{pages}{245132} (\bibinfo{year}{2011}).
	\newblock \urlprefix\url{https://journals.aps.org/prb/abstract/10.1103/PhysRevB.83.245132}.
	
\bibitem{Fang2013}
	\bibinfo{author}{Fang, C.}, \bibinfo{author}{Gilbert, M.~J.} \& \bibinfo{author}{Bernevig, B.~A.}
	\newblock \bibinfo{title}{Entanglement spectrum classification of $C_n$-invariant noninteracting topological insulators in two dimensions}.
	\newblock \emph{\bibinfo{journal}{Phys. Rev. B}} \textbf{\bibinfo{volume}{87}}, \bibinfo{pages}{035119} (\bibinfo{year}{2013}).
	\newblock \urlprefix\url{https://journals.aps.org/prb/abstract/10.1103/PhysRevB.87.035119}.


	
\bibitem{Bernevig2006}
	\bibinfo{author}{Bernevig, B.~A.}, \bibinfo{author}{Hughes, T.~L.} \& \bibinfo{author}{Zhang, S.-C.}
	\newblock \bibinfo{title}{Quantum Spin Hall Effect and Topological Insulator in HgTe Quantum Wells}.
	\newblock \emph{\bibinfo{journal}{Science}} \textbf{\bibinfo{volume}{314}}, \bibinfo{pages}{1757} (\bibinfo{year}{2006}).
	\newblock \urlprefix\url{https://www.science.org/doi/10.1126/science.1133734}.



\bibitem{Maiellaro2022}
	\bibinfo{author}{Maiellaro, A.} \emph{et~al.}
	\newblock \bibinfo{title}{Topological squashed entanglement: Nonlocal order parameter for one-dimensional topological superconductors}.
	\newblock \emph{\bibinfo{journal}{Phys. Rev. Research}} \textbf{\bibinfo{volume}{4}}, \bibinfo{pages}{033088} (\bibinfo{year}{2022}).
	\newblock \urlprefix\url{https://doi.org/10.1103/PhysRevResearch.4.033088}

\bibitem{Maiellaro2022b}
	\bibinfo{author}{Maiellaro, A.} \emph{et~al.}
	\newblock \bibinfo{title}{Edge states, Majorana fermions, and topological order in superconducting wires with generalized boundary conditions}.
	\newblock \emph{\bibinfo{journal}{Phys. Rev. B}} \textbf{\bibinfo{volume}{106}}, \bibinfo{pages}{155407} (\bibinfo{year}{2022}).
	\newblock \urlprefix\url{https://doi.org/10.1103/PhysRevB.106.155407}

\bibitem{Maiellaro2023}
	\bibinfo{author}{Maiellaro, A.} \emph{et~al.}
	\newblock \bibinfo{title}{Squashed entanglement in one-dimensional quantum matter}.
	\newblock \emph{\bibinfo{journal}{Phys. Rev. B}} \textbf{\bibinfo{volume}{107}}, \bibinfo{pages}{115160} (\bibinfo{year}{2023}).
	\newblock \urlprefix\url{https://doi.org/10.1103/PhysRevB.107.115160}

\bibitem{Maiellaro2023b}
     \bibinfo{author}{Maiellaro, A.} \emph{et~al.}
     \newblock \bibinfo{title}{Resilience of topological superconductivity under particle current}.
     \newblock \emph{\bibinfo{journal}{Phys. Rev. B}} \textbf{\bibinfo{volume}{107}}, \bibinfo{pages}{064505} (\bibinfo{year}{2023}).
     \newblock \urlprefix\url{https://doi.org/10.1103/PhysRevB.107.064505}
	

\bibitem{Liu2024}
	\bibinfo{author}{Liu, L.} \emph{et~al.}
	\newblock \bibinfo{title}{Engineering second-order topological insulators via coupling two first-order topological insulators}.
	\newblock \emph{\bibinfo{journal}{Phys. Rev. B}} \textbf{\bibinfo{volume}{110}}, \bibinfo{pages}{115427} (\bibinfo{year}{2024}).
	\newblock \urlprefix\url{https://journals.aps.org/prb/abstract/10.1103/PhysRevB.110.115427}.


\bibitem{Peschel2009}
	\bibinfo{author}{Peschel, I.} \& \bibinfo{author}{Eisler, V.}
	\newblock \bibinfo{title}{Reduced density matrices and entanglement entropy in free lattice models}.
	\newblock \emph{\bibinfo{journal}{J. Phys. A}} \textbf{\bibinfo{volume}{42}}, \bibinfo{pages}{504003} (\bibinfo{year}{2009}).
	\newblock \urlprefix\url{https://iopscience.iop.org/article/10.1088/1751-8113/42/50/504003}.
	
\bibitem{Peschel2003}
	\bibinfo{author}{Peschel, I.}
	\newblock \bibinfo{title}{Calculation of reduced density matrices from correlation functions}.
	\newblock \emph{\bibinfo{journal}{J. Phys. A}} \textbf{\bibinfo{volume}{36}}, \bibinfo{pages}{L205} (\bibinfo{year}{2003}).
	\newblock \urlprefix\url{https://iopscience.iop.org/article/10.1088/0305-4470/36/14/101}.
	
\bibitem{Arad2024}
	\bibinfo{author}{Arad, I.}, \bibinfo{author}{Firanko, R.} \& \bibinfo{author}{Jain, R.}
	\newblock \bibinfo{title}{An Area Law for the Maximally-Mixed Ground State in Arbitrarily Degenerate Systems with Good AGSP}.
	\newblock \emph{\bibinfo{journal}{Proc. 56th Annu. ACM Symp. Theory Comput.}} (\bibinfo{year}{2024}).
	\newblock \urlprefix\url{https://doi.org/10.1145/3618260.3649612}.


\end{thebibliography}
\end{document}